\documentclass[aps, superscriptaddress, floatfix]{revtex4}

\usepackage{graphicx}

\begin{document}



\newcommand{\Dirac}{\rlap{\hspace{-.5mm} \slash} D}
\newcommand{\sumint}{\rlap{\hspace{-.5mm} $\sum$} \int}
\title{
 The QCD phase diagram at nonzero baryon, isospin and strangeness
 chemical potentials: 
  Results from a hadron resonance gas model}

\author{D.~Toublan and John~B.~Kogut}
\affiliation {Physics Department, University of Illinois at Urbana-Champaign,
Urbana, IL 61801}

\date{\today}

\begin{abstract}
We use a hadron resonance gas model to study the
QCD phase diagram at nonzero temperature, baryon, isospin and strangeness
chemical potentials. We determine the temperature of the transition from
the hadronic phase to the quark gluon plasma phase using two different
methods.
We find that the critical temperatures derived in both methods
are in very good agreement.
We find that the critical surface has a small curvature.
We also find that the critical temperature's dependence on the
baryon chemical potential at zero isospin chemical
potential is almost identical to its dependence on the
isospin chemical potential at vanishing baryon chemical potential.
This result, which holds when the chemical potentials are small,
supports recent lattice simulation studies.
Finally, we find that at a given baryon
chemical potential, the critical temperature is lowered as either the
isospin or the strangeness
chemical potential are increased. Therefore, in order to lower the
critical temperature, it might be useful to use different isotopes in
heavy ion collision experiments.

\end{abstract}

\maketitle

\section{Introduction}
The QCD phase diagram at nonzero temperature and baryon density has been the
subject of numerous studies during the past few years \cite{cscRev,lattRev}.
Two regions of the
phase diagram have been placed under special scrutiny. First, in the low
temperature and high density domain, the ground state is believed to
be a color superconductor
\cite{cscOrig,cscMIT,cscStonyBrook,cscRev}. There are numerous
types of color
superconducting phases that emerge as the baryon chemical potential
increases \cite{cscRev,cscPhases}.
Second, in the high temperature and low baryon density
domain, there is a transition from the hadronic
phase to the quark gluon plasma phase \cite{lattRev}.
This latter domain is probed by heavy ion collision experiments.

In heavy ion collision experiments baryon number, isospin and strangeness
are conserved. The time between the formation of the fireball and its
freeze-out is so short that only the strong interactions play a
significant role.
In the past heavy ion collision experiments, 
the value of the strangeness chemical potential is $\sim 25\%$ of the
value of the baryon chemical potential, whereas the value of the 
isospin chemical potential is $\sim 2\%$ of the value of the baryon
chemical potential \cite{hicRev}.
Thus the strangeness and isospin chemical potentials are small but
not negligible. It is therefore  worthwhile to study the QCD phase
diagram in the high temperature region with nonzero baryon, isospin
and strangeness chemical potentials. 

At zero chemical potentials, numerical simulations on the lattice find
that the temperature of the transition from the hadronic phase to the
quark gluon plasma phase is given by $T_c=175 \pm 6$~MeV
\cite{lattT_Bielefeld,lattT_CP-PACS,lattT_F&K,lattT_MILC}.
However, at nonzero baryon chemical
potential lattice simulations
suffer from the so-called sign problem: The fermion
determinant is complex. As a consequence,
traditional methods cannot be used to study the QCD phase diagram at
nonzero baryon chemical potential. However, recent advances have
allowed studies of the high temperature and low baryon chemical
potential region
\cite{lattMuB_F&K,lattMuB_Bielefeld,lattMuB_ZH,lattMuB_Maria,
lattMuB_Gupta,lattMuB_Azcoiti}.
The situation at zero baryon and strangeness chemical
potentials and nonzero isospin chemical potential is simpler: There is
no sign problem and traditional methods can be used
\cite{lattMuI} (as in QCD with two colors at nonzero
baryon chemical potential \cite{QCD2}).
The results in this case are in complete agreement with effective theory
studies \cite{chptMu}. From these lattice studies it appears that the
critical temperature's dependence on small baryon chemical potential
at zero isospin and strangeness chemical
potentials matches its dependence on small isospin chemical potential at
vanishing baryon and strangeness chemical potentials
\cite{lattMuB_F&K,lattMuB_Bielefeld,lattMuB_ZH,lattMuI}.
We will show that the hadron resonance gas model supports this
conclusion.

Recently, the study of the QCD phase diagram at nonzero temperature,
baryon and isospin chemical potentials has attracted attention
\cite{qcdMuBMuI_RMT,qcdMuBMuI_NJL,qcdMuBMuI_Ladder}.
Using different models, it was found that an
arbitrarily small isospin chemical potential could greatly alter the
QCD phase diagram at high temperature and small baryon chemical
potential and have important consequences for heavy ion collision
experiments. There are two phase transitions at high
temperature. There
are phases where the $u$ and $d$ quark sectors are decoupled
\cite{qcdMuBMuI_RMT,qcdMuBMuI_NJL,qcdMuBMuI_Ladder}.
These results need to be confirmed
by other methods, in particular on the lattice.

We study the QCD phase diagram at high
temperature and nonzero baryon, isospin and strangeness chemical
potentials using the hadron resonance gas model. It has been shown
both experimentally and 
on the lattice that the hadronic phase is very well described by a
weakly interacting hadron resonance gas \cite{hicRev,hgrExp,hgrLatt}.
We use two different methods to determine
the location of the transition. First,
it has been
found on the lattice that the phase transition that separates the
hadronic phase from
the quark gluon plasma phase corresponds to a surface of constant energy
density: $\epsilon \sim 0.5 - 1.0$~GeV/fm$^3$ \cite{hgrLatt}.
Second, the quark-antiquark condensate for the $u$ and $d$
quarks should almost vanish at the transition temperature.
In this article, we determine the critical temperature, $T_c$, at nonzero
baryon, isospin
and strangeness chemical potentials by using both approaches. We
compute the surfaces of constant energy density as well
as the quark-antiquark condensate in a
hadron resonance gas  model at nonzero temperature, baryon,
isospin and strangeness
chemical potentials. We show that both methods agree qualitatively as
well as quantitatively.  We find that the
critical surface has small
curvature, and that the critical temperature slowly decreases when either the
baryon, the isospin, or the strangeness chemical potentials are
increased.

\section{The model}

We assume that the pressure in the hadronic phase
is given by the contributions of all the
hadron resonances up to $2$~GeV treated as a free gas, as in
\cite{hgrLatt}. All the
thermodynamic observables can be derived from the pressure since
\begin{eqnarray}
  \label{pZ}
  p = \lim_{V\rightarrow\infty} \, \frac TV \, \ln \,
  Z(T,\mu_B,\mu_I,\mu_S, V),
\end{eqnarray}
where $ Z(T,\mu_B,\mu_I,\mu_S,V)$ is the grand canonical partition
function in a finite volume $V$, at nonzero temperature, $T$, baryon
chemical potential, $\mu_B$,  isospin
chemical potential, $\mu_I$, and strangeness chemical potential,
$\mu_S$. The energy
density is given by
\begin{eqnarray}
  \label{enZ}
  \epsilon = T \frac{\partial p}{\partial T}-p+\mu_B \frac{\partial
    p}{\partial \mu_B} 
  +\mu_I \frac{\partial p}{\partial \mu_I} 
  +\mu_S \frac{\partial p}{\partial \mu_S} .
\end{eqnarray}
For a quark $q$ with mass $m_q$, the quark-antiquark condensate is given by
\begin{eqnarray}
  \label{qqZ}
  \langle \bar{q} q \rangle =  \frac{\partial p}{\partial m_q}.
\end{eqnarray}

At nonzero temperature, the contributions of massive states are
exponentially suppressed $\sim \exp(-m_{\rm H}/T)$. Their
interactions are also exponentially suppressed $\sim \exp(-(m_{\rm
H}+m'_{\rm H})/T)$. Therefore this approximation should be valid at
low enough temperatures. However, since we are studying QCD at
temperatures up to $\sim200$~MeV, the lightness of the pions could
be a problem, since $m_\pi\simeq140$~MeV. The hadron resonance gas
model should be a good approximation for the other hadrons since
they have a mass larger than $\sim$ $500$~MeV. The physics of pions
at nonzero temperature has been extensively studied in chiral
perturbation theory \cite{chptGL,chptT}. The pions' contributions to
the pressure have been calculated up to three loops in chiral
perturbation theory \cite{chptGL}. It has been shown that the free
gas approximation and chiral perturbation theory agree at the one
loop level, and that, in chiral perturbation theory, the two-loop
corrections to the pressure are below a few percents of the one-loop
contributions for temperatures under $200$~MeV \cite{chptGL}. Thus
the hadron gas resonance model is a good approximation also for the
pions. The hadron resonance gas model has already been used in the
literature and has been shown to give a very good description of the
hadronic phase and of the critical temperature \cite{hgrLatt}.

In the free gas approximation,
the contribution to the pressure due to a particle of mass $m_{\rm
  H}$, baryon charge $B$, isospin $I_3$, strangeness $S$, 
and degeneracy $g$ is given by
\begin{eqnarray}
  \label{p1}
  \Delta p=g m_{\rm H}^2 T^2 \,
\sum_{n=1}^\infty \,\frac{(-\eta)^{n+1}}{2 n^2 \pi^2}
\,\exp\left( \frac{n (B\mu_B - I_3\mu_I - S \mu_S)}{T} \right)
\, K_2\left( \frac{n m_{\rm H}}{T} \right),
\end{eqnarray}
where $\eta=+1$ for fermions
and $\eta=-1$ for bosons, and $K_n(x)$ is the modified Bessel function.
This particle's contribution to the energy density is given
by
\begin{eqnarray}
  \label{en1}
  \Delta \epsilon=g m_{\rm H}^2 T \,
\sum_{n=1}^\infty \,\frac{(-\eta)^{n+1}}{2 n^2 \pi^2}
\,\exp\left( \frac{n(B\mu_B - I_3\mu_I - S \mu_S)}{T} \right)
\,  \left[ 3 T K_2\left( \frac{n m_{\rm H}}{T} \right)
+ n m_{\rm H}  K_1\left( \frac{n m_{\rm H}}{T} \right)
\right],
\end{eqnarray}
and its contribution to the quark-antiquark
condensate is given by
\begin{eqnarray}
  \label{qq1}
  \Delta \langle\bar{q}q\rangle = -g m_{\rm H}^2 T \,
\frac{\partial m_{\rm H}}{\partial m_q} \,
\sum_{n=1}^\infty \,\frac{(-\eta)^{n+1}}{2 n \pi^2}
\,\exp\left( \frac{n (B\mu_B - I_3\mu_I - S \mu_S)}{T} \right)
\, K_1\left( \frac{n m_{\rm H}}{T} \right).
\end{eqnarray}
In order to compute (\ref{qq1}), we need to know $m_{\rm H}$ as a
function of $m_u$ or $m_d$.
We make two assumptions in order to compute (\ref{qq1}).
First, we assume that the Gell-Mann$-$Oakes$-$Renner relation is valid
\begin{eqnarray}
  \label{GOR}
  F_\pi^2  \, m_\pi^2=(m_u+m_d) \langle\bar{q}q\rangle_0,
\end{eqnarray}
where $F_\pi=93$~MeV is the pion decay constant, and
$\langle\bar{q}q\rangle_0=\langle\bar{u}u\rangle_0=\langle\bar{d}d\rangle_0$
is the quark-antiquark condensate at zero
temperature and chemical potentials.
Second, based on lattice results, we assume that the pion mass
dependence of the hadron masses is given by
\begin{eqnarray}
  \label{mHmPi}
  \frac{\partial m_{\rm H}}{\partial (m_\pi^2)} \simeq
  \frac{A}{m_{\rm H}},
\end{eqnarray}
where $0.9 \lesssim A \lesssim 1.2$ \cite{hgrLatt}.
Therefore, combining (\ref{GOR}) and (\ref{mHmPi}), we assume that
\begin{eqnarray}
  \label{mHmQ}
  \frac{\partial m_{\rm H}}{\partial m_q} \simeq
  \frac{A \langle\bar{q}q\rangle_0}{ F_\pi^2 \, m_{\rm H}}.
\end{eqnarray}
Notice that since the hadron spectrum is isospin symmetric, we have
that
$\langle\bar{u}u\rangle=\langle\bar{d}d\rangle\equiv\langle\bar{q}q\rangle$
in the hadron resonance model at any temperature, baryon, isospin and
strangeness chemical
potentials. Therefore, the rich structure of the phase diagram found in
\cite{qcdMuBMuI_RMT,qcdMuBMuI_NJL,qcdMuBMuI_Ladder} cannot be seen in
this model. Finally at fixed
$T$, it can be readily seen from (\ref{en1}) and (\ref{qq1}) that an
increase in either $\mu_B$, or $\mu_I$, or $\mu_S$ will increase
$\epsilon$  and decrease $\langle\bar{q}q\rangle$.
Thus using either $\epsilon$ or
$\langle\bar{q}q\rangle$ as a criterion to determine the critical
temperature,we find that an increase in either $\mu_B$, or $\mu_I$, or
$\mu_S$ decreases
$T_c$, and that at fixed $\mu_B$ an increase in $\mu_I$ or $\mu_S$
 results in a decrease of $T_c$ as well.

\section{Results}

\subsection{Energy density criterion}
\begin{figure}[h]
\hspace{-0.8cm}
\includegraphics*[scale=0.50, clip=true, angle=0,
draft=false]{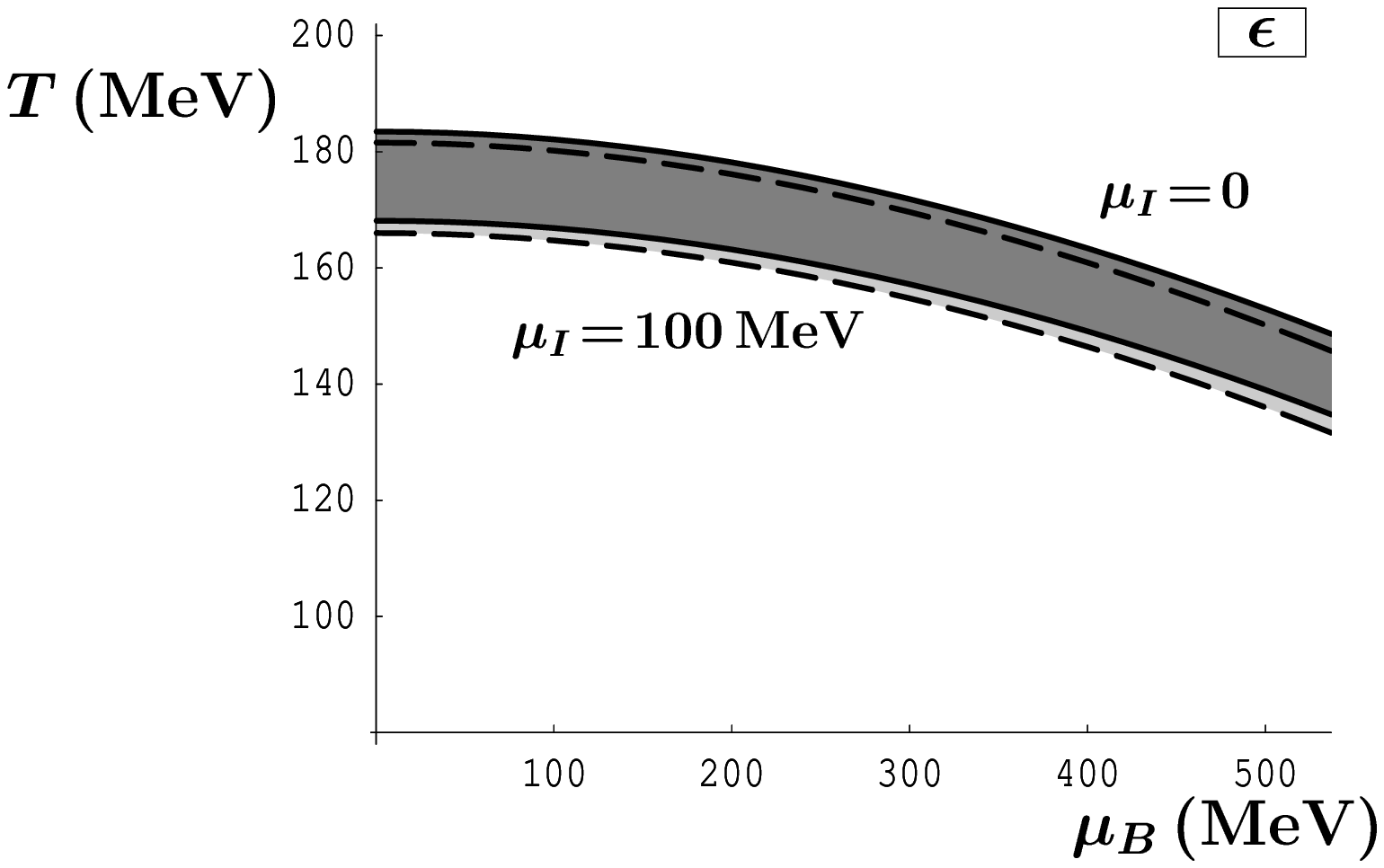}
\hspace{1.2cm}
\includegraphics*[scale=0.50, clip=true, angle=0,
draft=false]{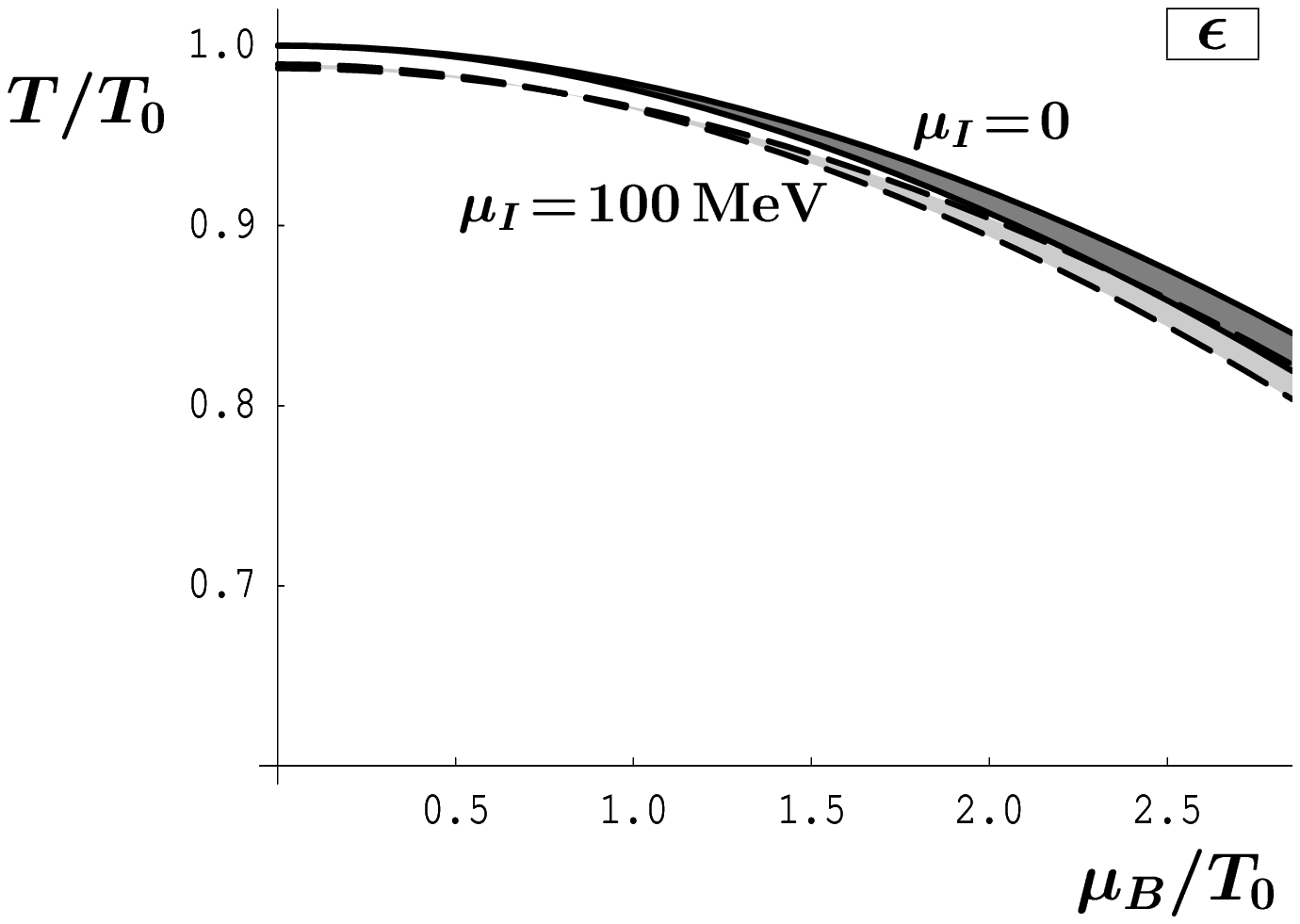}\\
\vspace{1cm}
\hspace{-0.8cm}
\includegraphics*[scale=0.50, clip=true, angle=0,
draft=false]{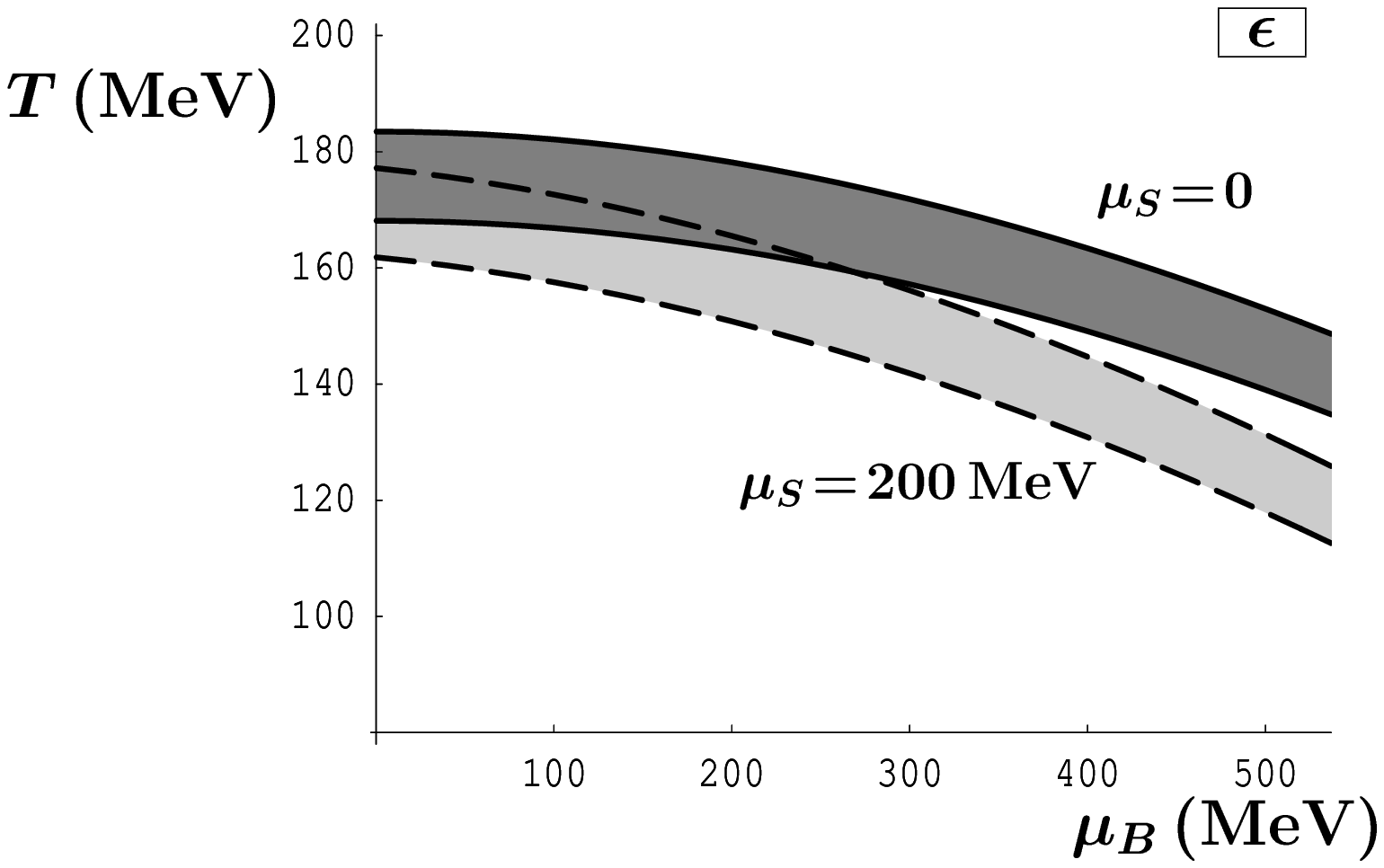}
\hspace{1.2cm}
\includegraphics*[scale=0.50, clip=true, angle=0,
draft=false]{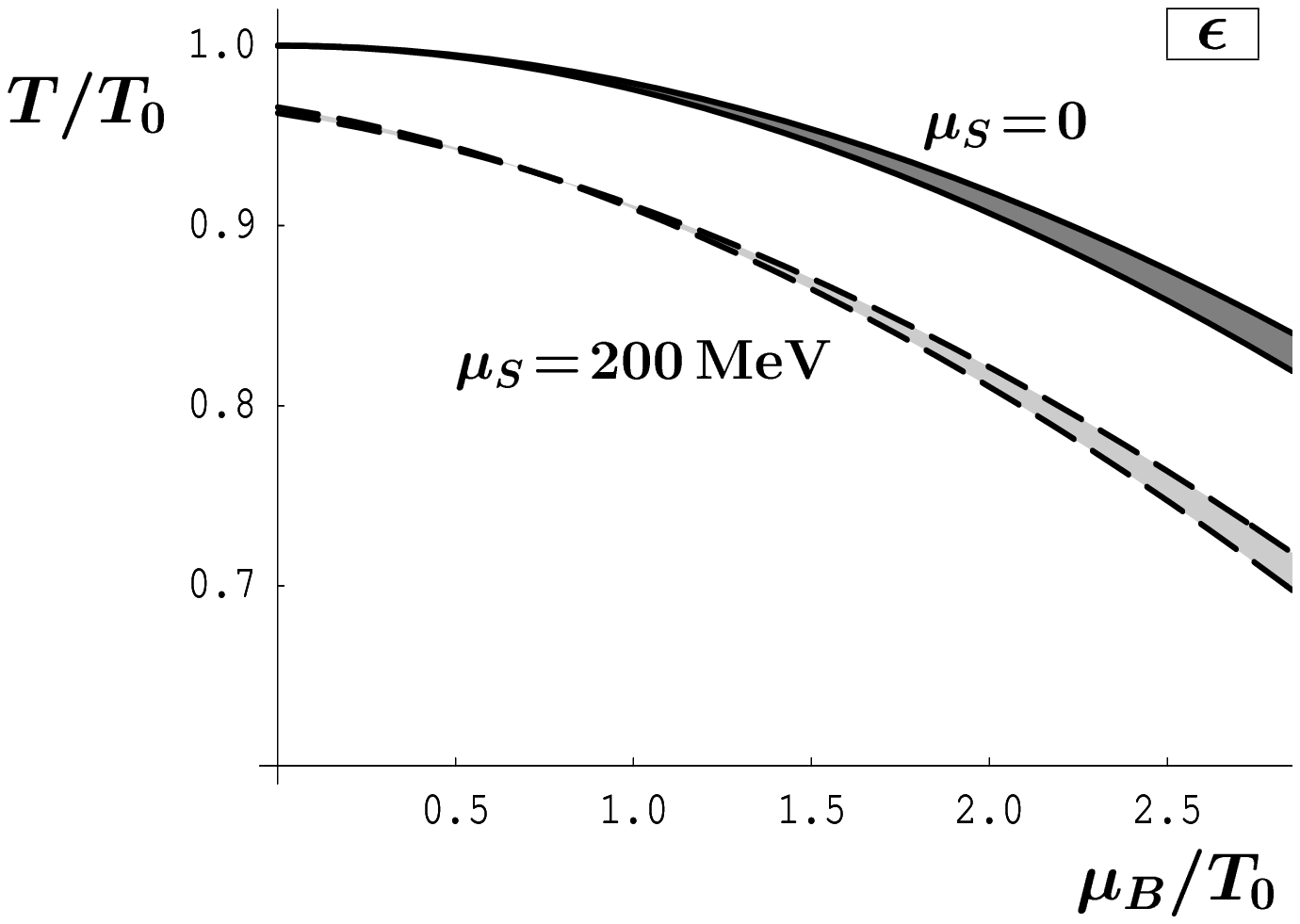}
\caption{\label{fig1} Critical temperature as a function of
$\mu_B$ determined by lines of constant energy
  density: $\epsilon \simeq 0.5 - 1.0 $~GeV/fm$^3$. 
  In the upper two plots $\mu_S=0$, the dark shading with full curves
  corresponds to $\mu_I=0$, and the light shading with dashed curves
  corresponds to $\mu_I=100$~MeV. 
  In the lower two plots $\mu_I=0$, the dark shading with full curves
  corresponds to $\mu_S=0$, and the light shading with dashed curves
  corresponds to $\mu_S=200$~MeV. 
  $T_0$ is the critical temperature at zero chemical potentials.}
\end{figure}

Lattice simulations have shown that the transition from the hadronic
phase to the quark gluon plasma phase takes place at a constant energy
density $\epsilon \simeq 0.5 - 1.0$~GeV/fm$^3$ \cite{hgrLatt}.
We use this criterion to determine the critical temperature as a
function of baryon, isospin and strangeness chemical potentials. Our
results for the critical temperature as a function of $\mu_B$ at
fixed $\mu_I$ and $\mu_S$ are shown in Figure~\ref{fig1}.
We find that this criterion constrains
the critical temperature in a band of $\sim15$~MeV.
At zero chemical potentials, we find that $T_c=176 \pm 8$~MeV, which is
in good agreement with lattice simulations
\cite{lattT_Bielefeld,lattT_CP-PACS,lattT_F&K,lattT_MILC}.
The temperature decreases as $\mu_B$ increases, as expected, but the
decrease is slow.  
At the accuracy we can achieve using this method, an increase in
$\mu_I$ does indeed decrease the critical
temperature at fixed $\mu_B$, but this effect is small at best.
The decrease of the critical temperature is more important when
$\mu_S$ is increased.
\begin{figure}[h]
\hspace{-0.8cm}
\includegraphics*[scale=0.50, clip=true, angle=0,
draft=false]{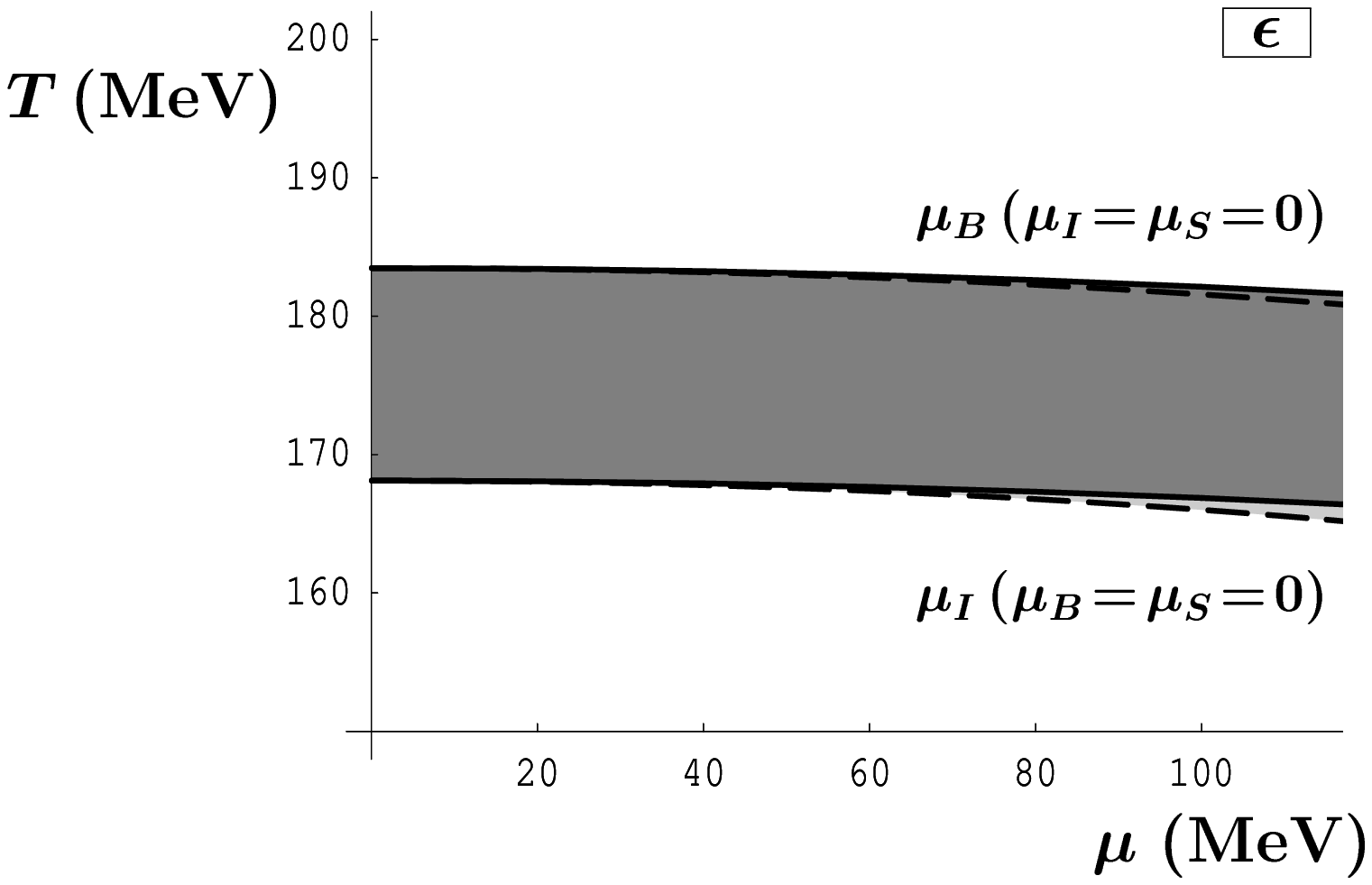}
\hspace{1.2cm}
\includegraphics*[scale=0.50, clip=true, angle=0,
draft=false]{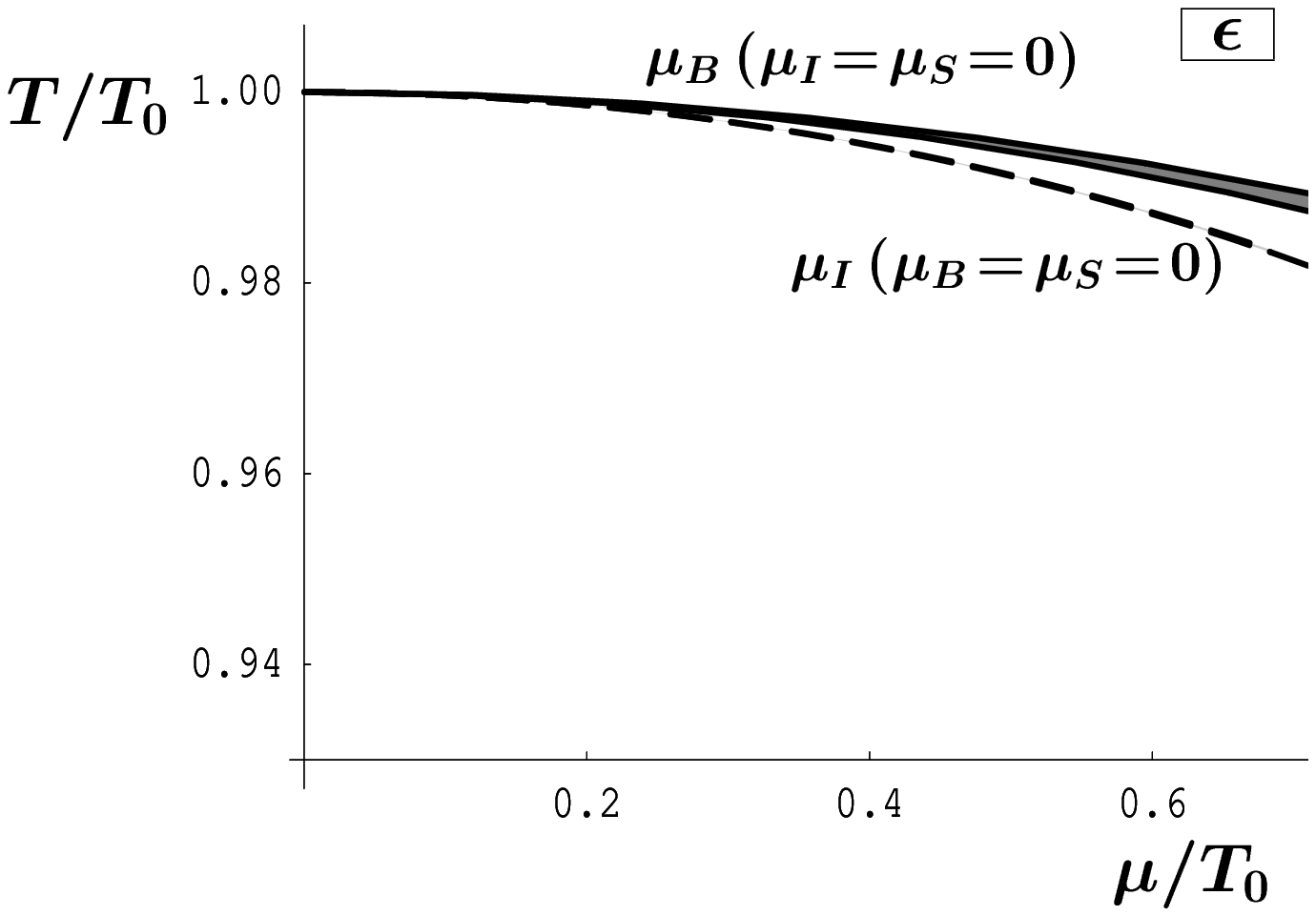}
\caption{\label{fig2} Critical temperature as a function of
$\mu_B$ at $\mu_I=\mu_S=0$ (dark shading
  with full curves), and as a function of
  $\mu_I$ at $\mu_B=\mu_S=0$ (light shading with
  dashed curves),
  determined by lines of constant energy
  density: $\epsilon \simeq 0.5 - 1.0$~GeV/fm$^3$.
  $T_0$ is the critical temperature at zero chemical potentials.}
\end{figure}
In Figure~\ref{fig2}, we compare the critical temperature
as a function of $\mu_B$ at $\mu_I=\mu_S=0$ with the critical temperature
as a function of $\mu_I$ at $\mu_B=\mu_S=0$. Notice that we limit ourselves to
$\mu_I \lesssim m_\pi$ and $\mu_S \lesssim m_K$ 
in order to avoid the pion and kaon superfluid phases
\cite{lattMuI,chptMu,kaonCond}.
We find that the critical temperature curves are almost identical in both
cases. This is in agreement with
results from the lattice
\cite{lattMuB_F&K,lattMuB_Bielefeld,lattMuB_ZH,lattMuB_Maria,lattMuI}.

We can fit our result for the critical temperature as a
function of $\mu_B$, $\mu_I$ and $\mu_S$. By construction, since the pressure
is an even function of $\mu_I$ in the hadron
resonance model, the critical
temperature is also even in $\mu_I$. We find
\begin{eqnarray}
  \label{critTenFit}
  \frac{T_c}{T_0}=1 - 0.021(2) \left( \frac{\mu_B}{T_0} \right)^2
- 0.039(1) \left( \frac{\mu_I}{T_0} \right)^2 
- 0.037(2) \left( \frac{\mu_S}{T_0} \right)^2 
- 0.031(3)  \; \frac{\mu_B \; \mu_S}{T_0^2} + \cdots,
\end{eqnarray}
where $T_0$ is the critical temperature at zero chemical potentials.
The fit is excellent, with a
linear regression coefficient $R^2=0.994$. 


\subsection{Quark-antiquark condensate criterion}
\begin{figure}[h]
\hspace{-0.8cm}
\includegraphics*[scale=0.50, clip=true, angle=0,
draft=false]{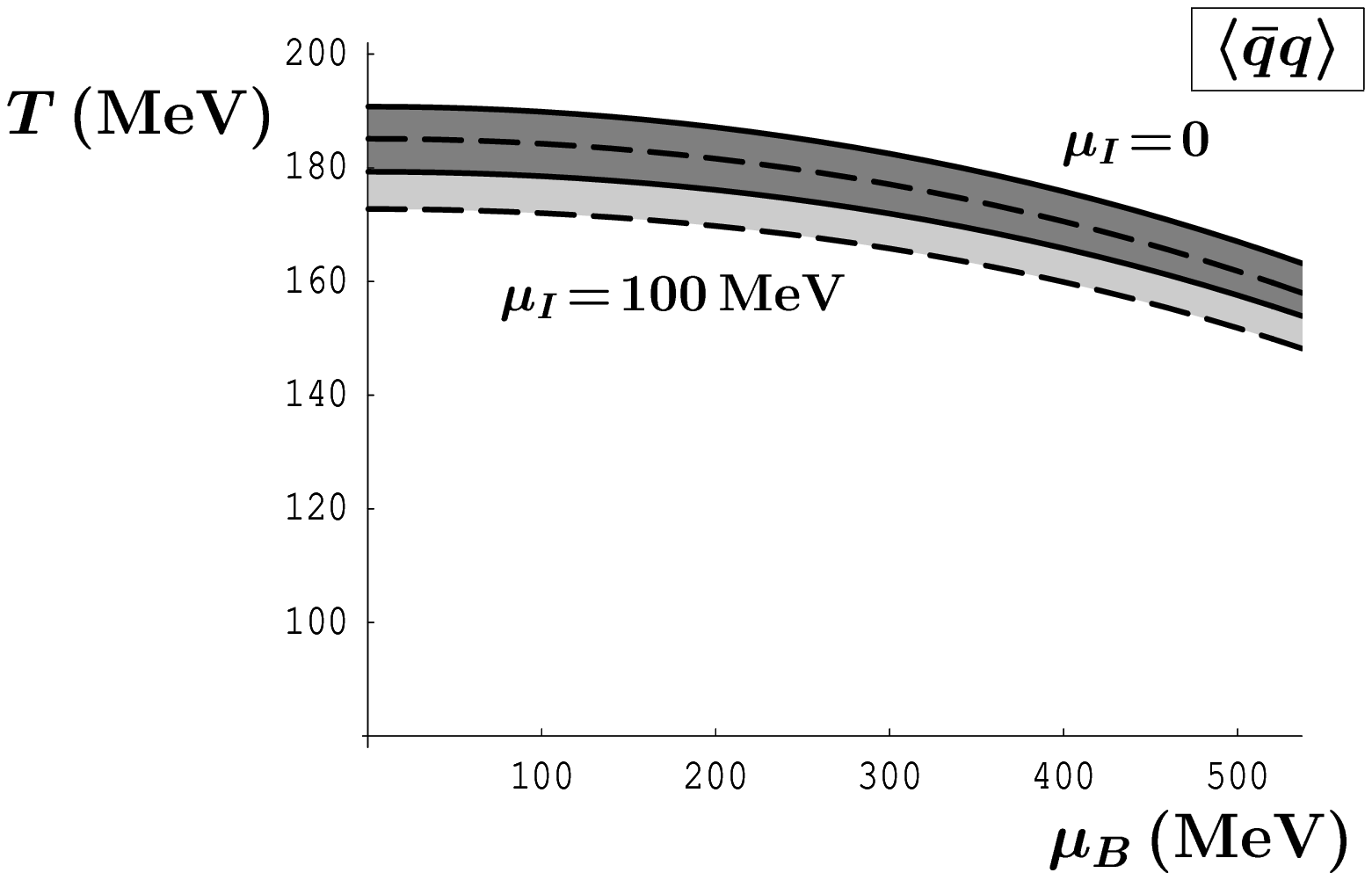}
\hspace{1.2cm}
\includegraphics*[scale=0.50, clip=true, angle=0,
draft=false]{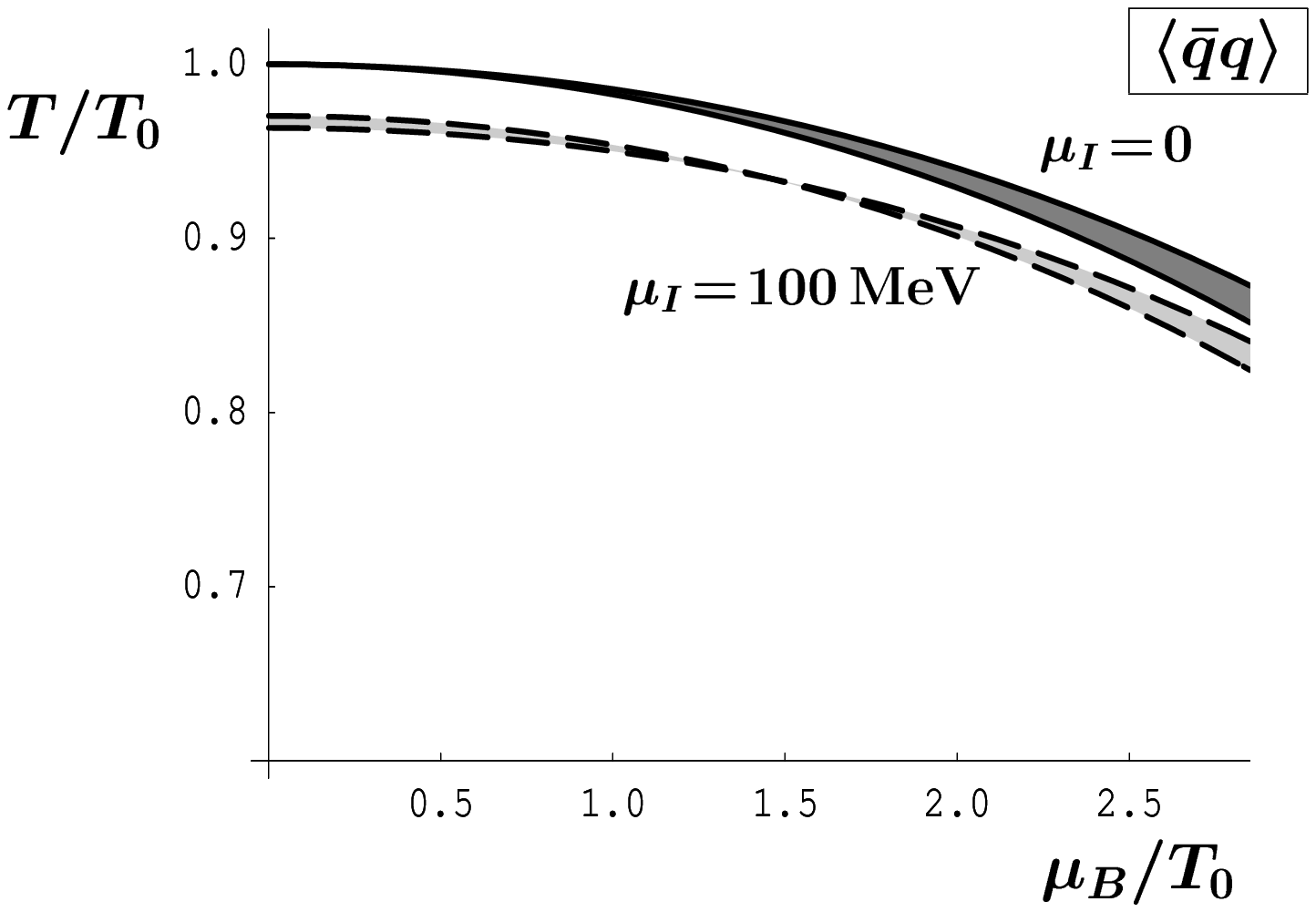}\\
\vspace{1cm}
\hspace{-0.8cm}
\includegraphics*[scale=0.50, clip=true, angle=0,
draft=false]{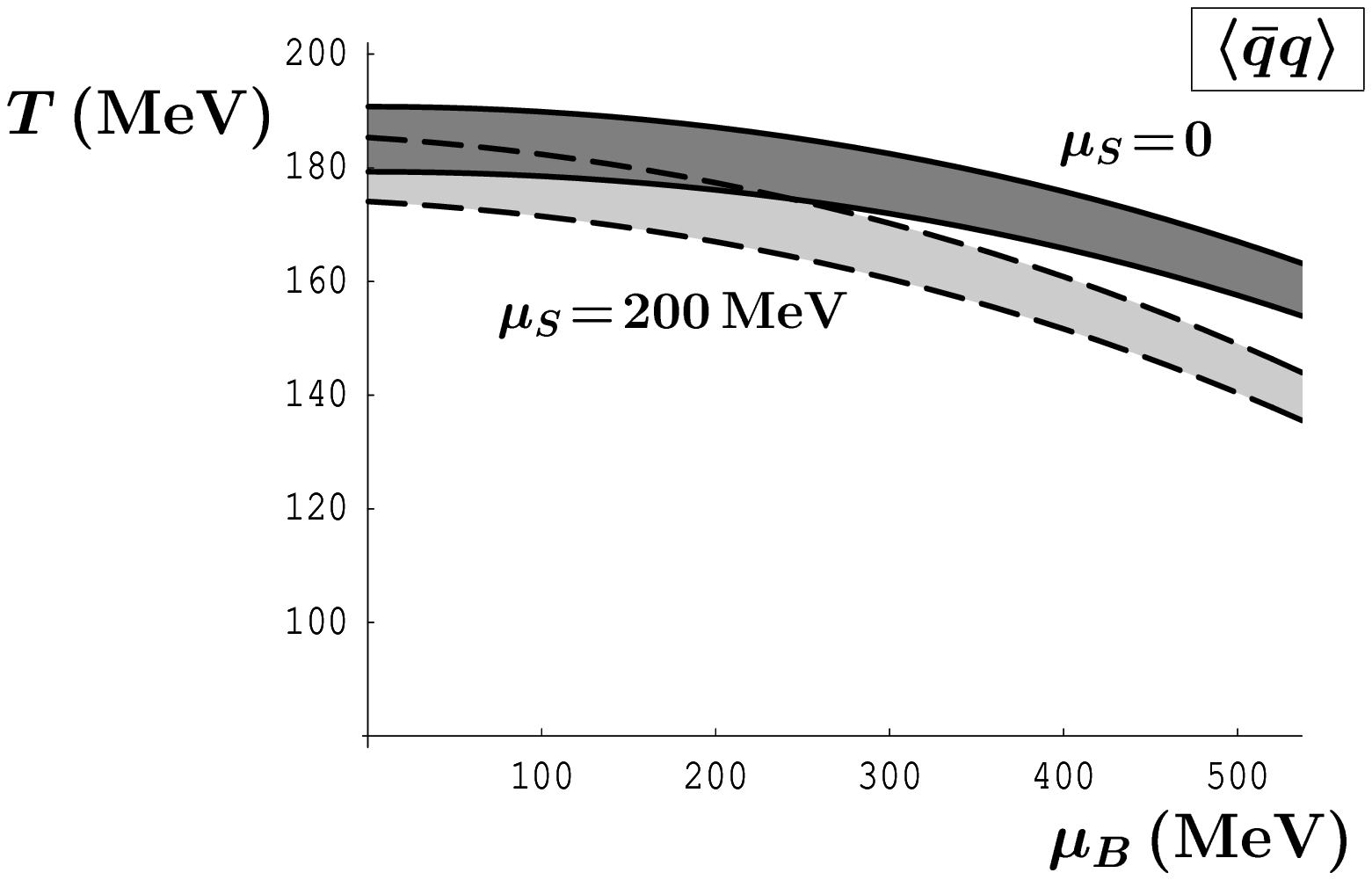}
\hspace{1.2cm}
\includegraphics*[scale=0.50, clip=true, angle=0,
draft=false]{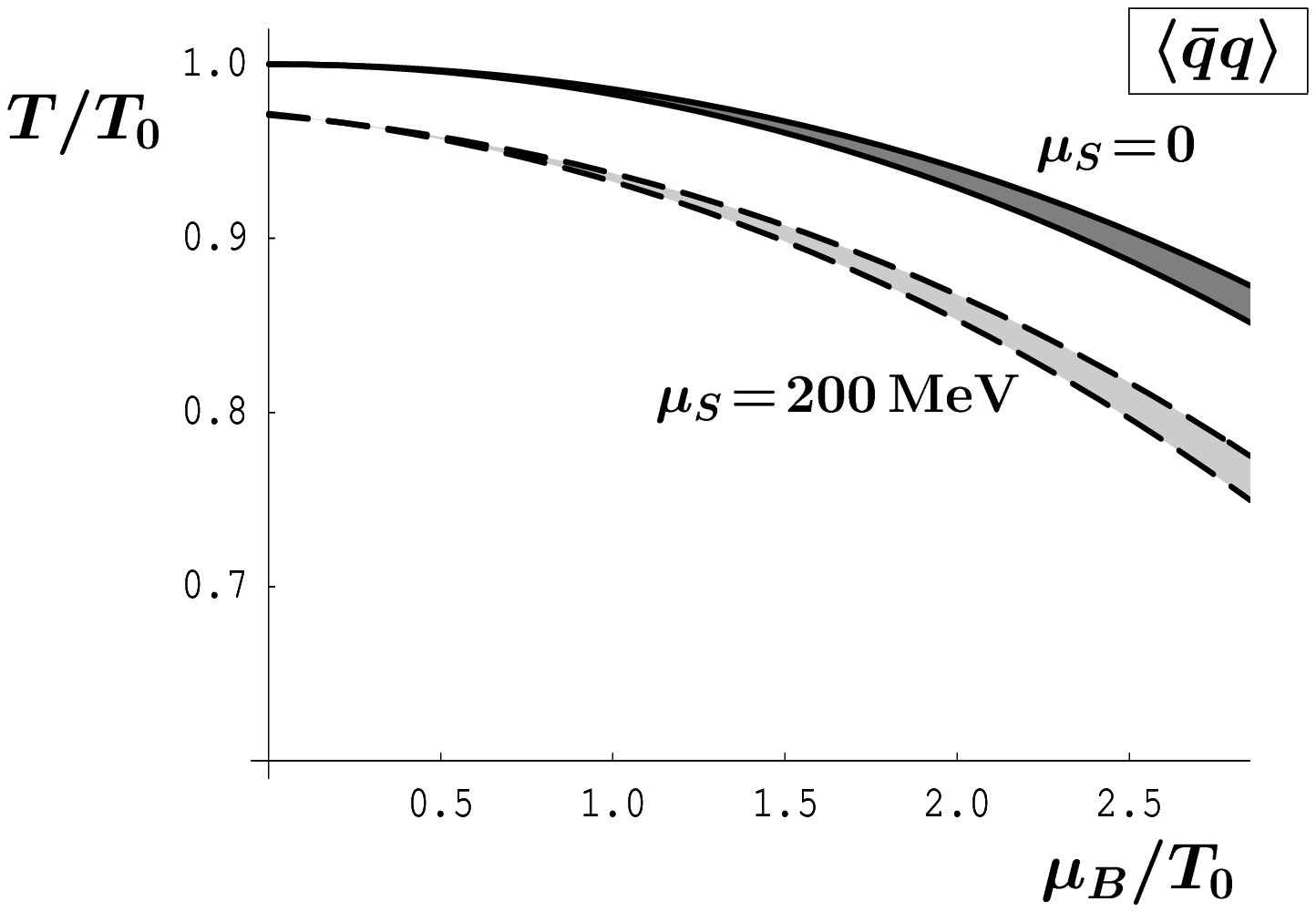}
\caption{\label{fig3} Critical temperature as a function of $\mu_q$
  determined by $\langle\bar{q}q\rangle=0$.
  In the upper two plots $\mu_S=0$, the dark shading with full curves
  corresponds to $\mu_I=0$, and the light shading with dashed curves
  corresponds to $\mu_I=100$~MeV. 
  In the lower two plots $\mu_I=0$, the dark shading with full curves
  corresponds to $\mu_S=0$, and the light shading with dashed curves
  corresponds to $\mu_S=200$~MeV. 
  $T_0$ is the critical temperature at zero chemical potentials.}
\end{figure}

\begin{figure}[h]
\hspace{-0.8cm}
\includegraphics*[scale=0.50, clip=true, angle=0,
draft=false]{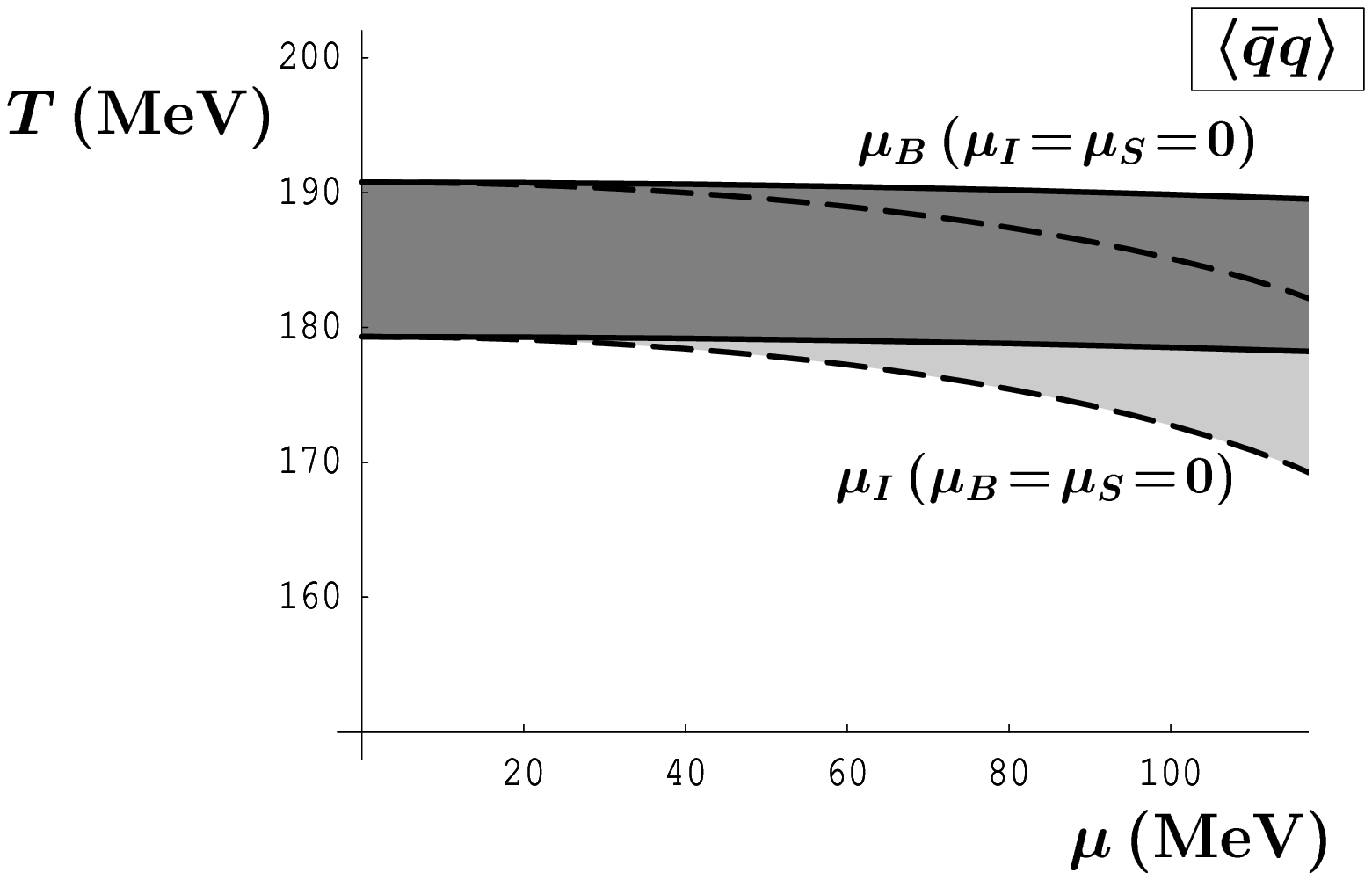}
\hspace{1.2cm}
\includegraphics*[scale=0.50, clip=true, angle=0,
draft=false]{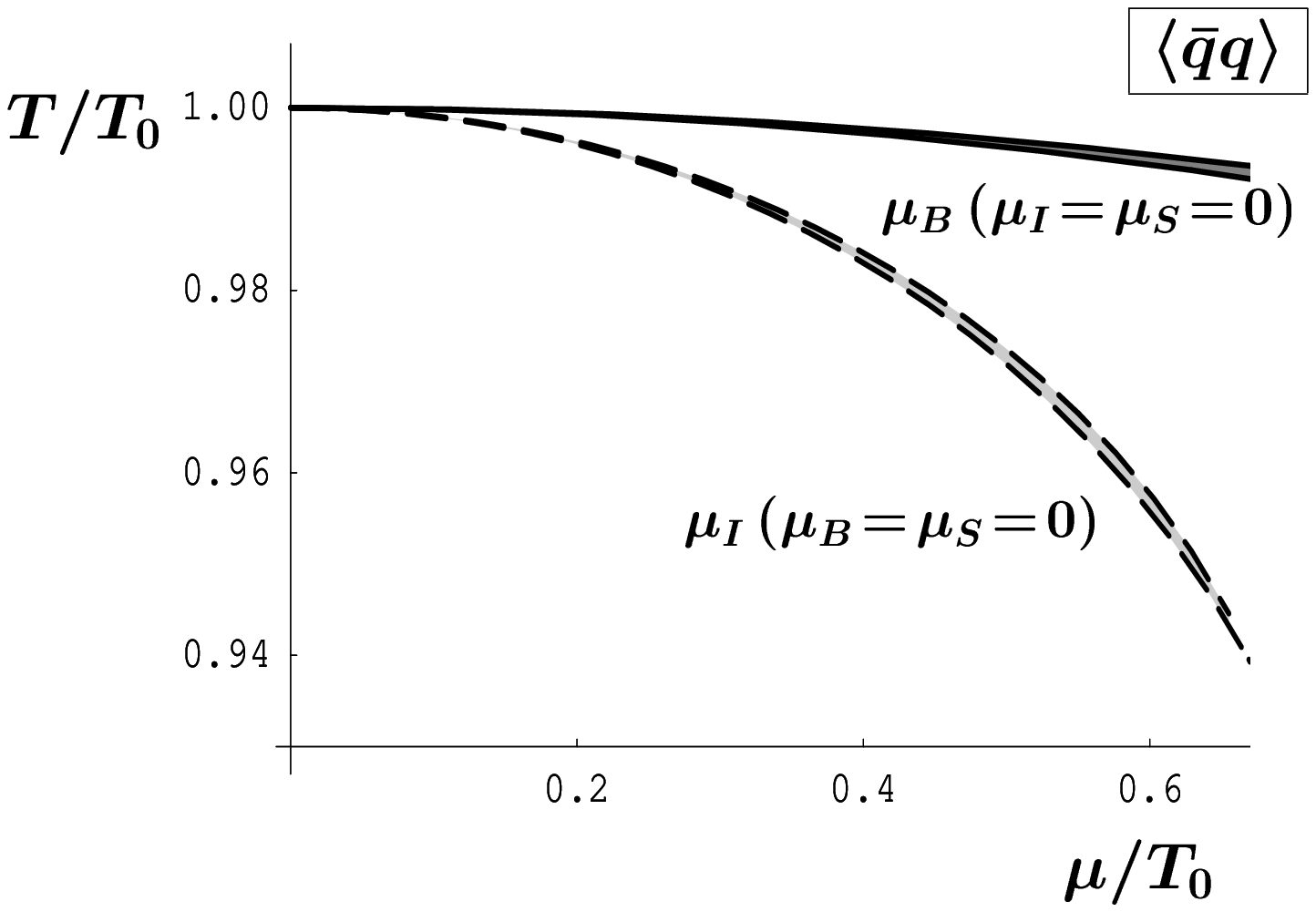}
\caption{\label{fig4} Critical temperature as a function of
$\mu_B$ at $\mu_I=\mu_S=0$ (dark shading
  with full curves), and as a function of
  $\mu_I$ at $\mu_B=\mu_S=0$ (light shading with
  dashed curves),
  determined by   $\langle\bar{q}q\rangle=0$.
  $T_0$ is the critical temperature at zero chemical potentials.}
\end{figure}

The critical temperature can also be computed from the quark-antiquark
condensate. Indeed $\langle\bar{q}q\rangle$ is of the order of the
light quark masses at the phase transition and therefore almost
vanishes.
We determine the critical
temperature by finding the point where
$\langle\bar{q}q\rangle=0$
in the hadron resonance gas model.
We obtain a range of critical temperatures, since in the
relation (\ref{mHmQ}) the constant $A \simeq 0.9 - 1.2$ is not
precisely known \cite{hgrLatt}.
Our results for the critical temperature as a function of $\mu_B$ at
fixed $\mu_I$ and $\mu_S$ using $\langle\bar{q}q\rangle$
are shown in Figure~\ref{fig3}.
We find that this criterion  constrains
the critical temperature to a band of $\sim 15$~MeV.
At zero chemical potentials, we find that $T_c=185 \pm 6$~MeV, which is
in good agreement both with the result obtained above using the
$\epsilon$-criterion, as well as  with lattice simulations
\cite{lattT_Bielefeld,lattT_CP-PACS,lattT_F&K,lattT_MILC}.
As expected, the critical
temperature decreases as $\mu_B$ increases. As in the previous method,
we find that an increase in the isospin chemical potential
might reduce the critical temperature, but not in a significant
way. The strangeness chemical potential has a stronger effect on the
critical temperature.
In Figure~\ref{fig4}, we compare the critical temperature
as a function of $\mu_B$ at $\mu_I=\mu_S=0$ with the critical temperature as
a function of $\mu_I$ at $\mu_B=\mu_S=0$.
We find that the critical temperature curves are almost identical in both
cases.

We can fit our result for the critical temperature as a
function of $\mu_B$, $\mu_I$ and $\mu_S$. We find
\begin{eqnarray}
  \label{critTsigmaFit}
  \frac{T_c}{T_0}=1 - 0.017(1) \left( \frac{\mu_B}{T_0} \right)^2
- 0.109(4) \left( \frac{\mu_I}{T_0} \right)^2 
- 0.032(2) \left( \frac{\mu_S}{T_0} \right)^2 
- 0.024(2)  \; \frac{\mu_B \; \mu_S}{T_0^2} + \cdots,
\end{eqnarray}
where $T_0$ is the critical temperature at zero chemical potentials.
The fit is excellent, with a
linear regression coefficient $R^2=0.991$. 

\begin{figure}[h]

\hspace{-0.8cm}
\includegraphics*[scale=0.50, clip=true, angle=0,
draft=false]{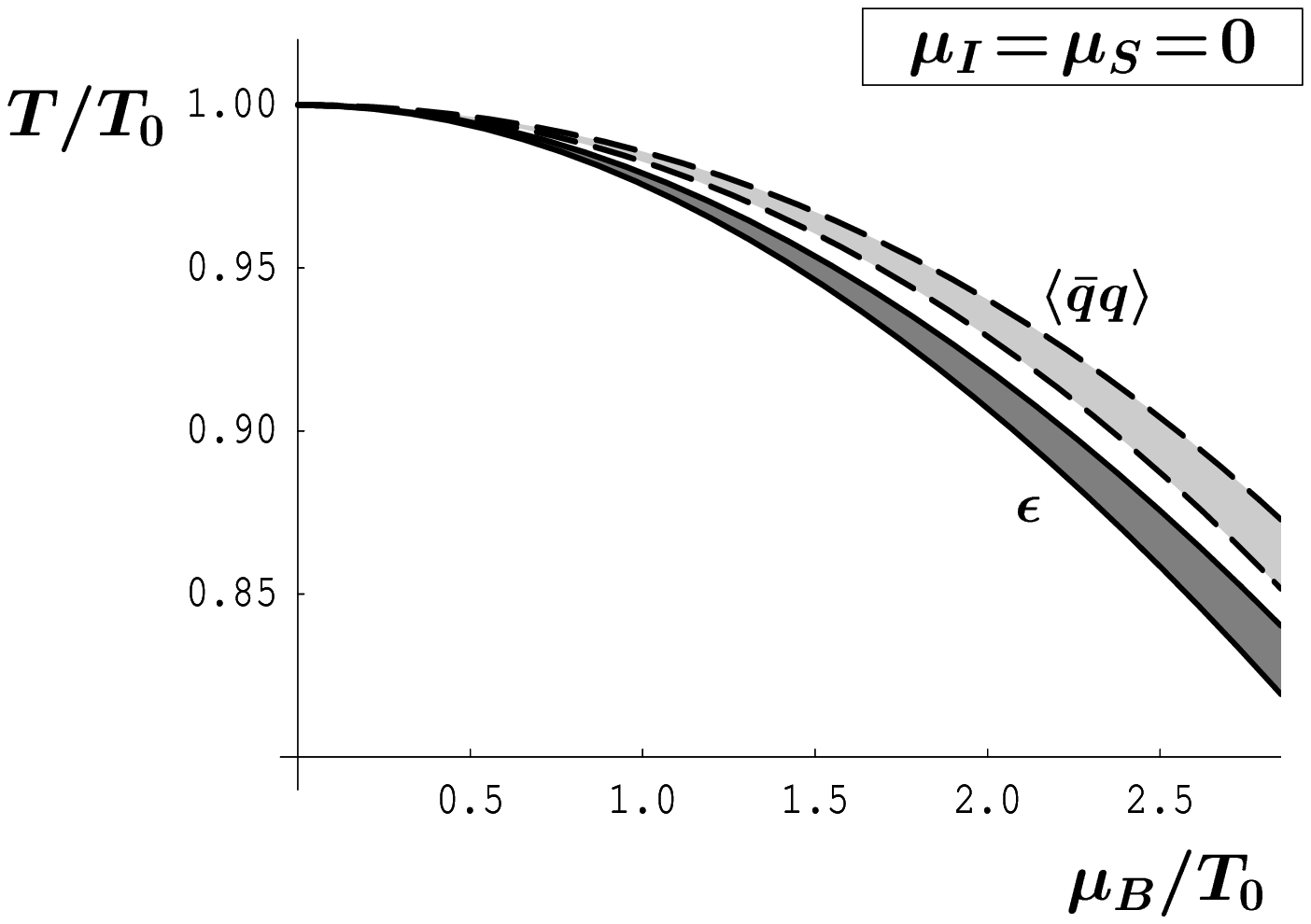}\\
\vspace{1cm}
\hspace{-0.8cm}
\includegraphics*[scale=0.50, clip=true, angle=0,
draft=false]{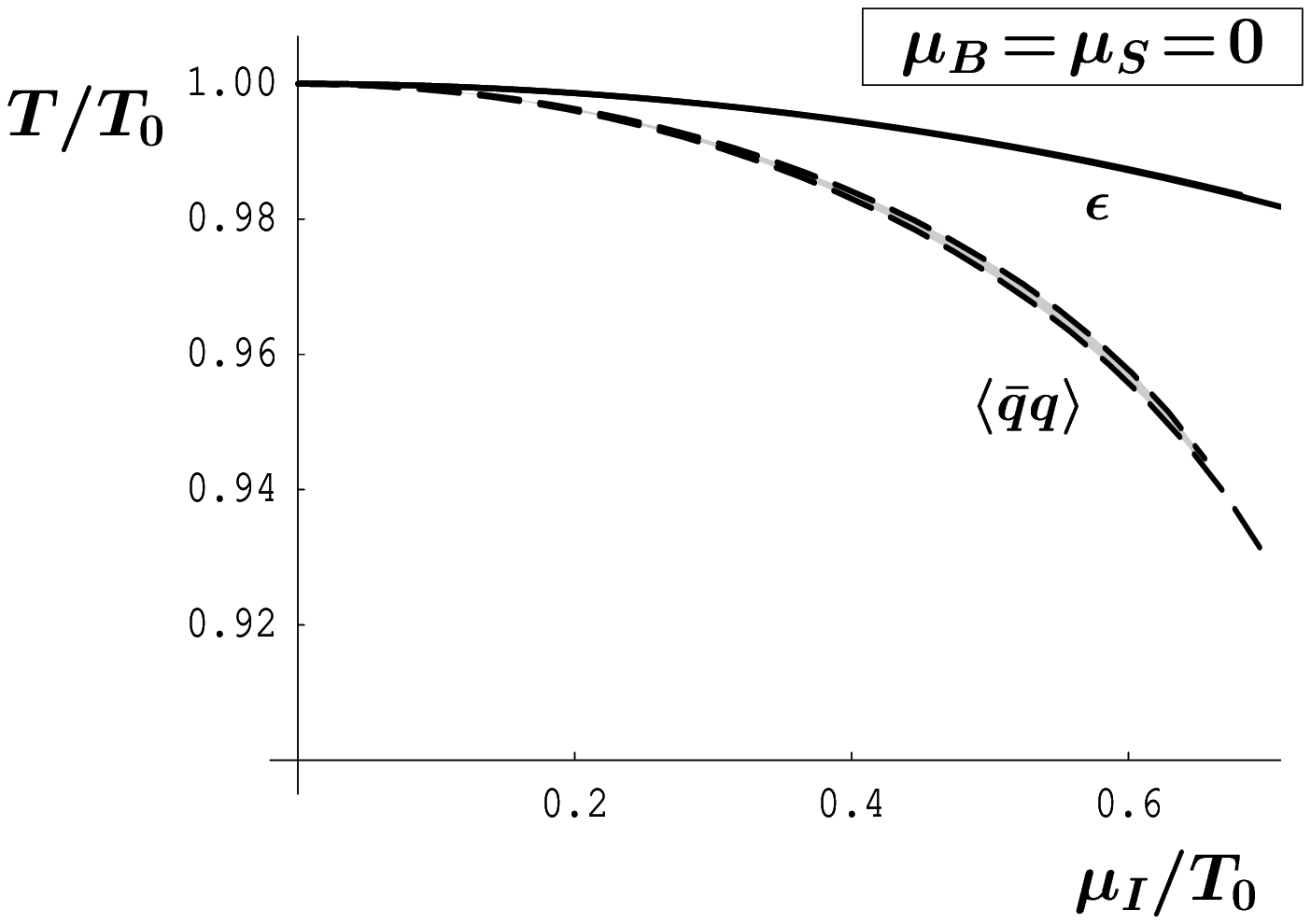}
\hspace{1.2cm}
\includegraphics*[scale=0.50, clip=true, angle=0,
draft=false]{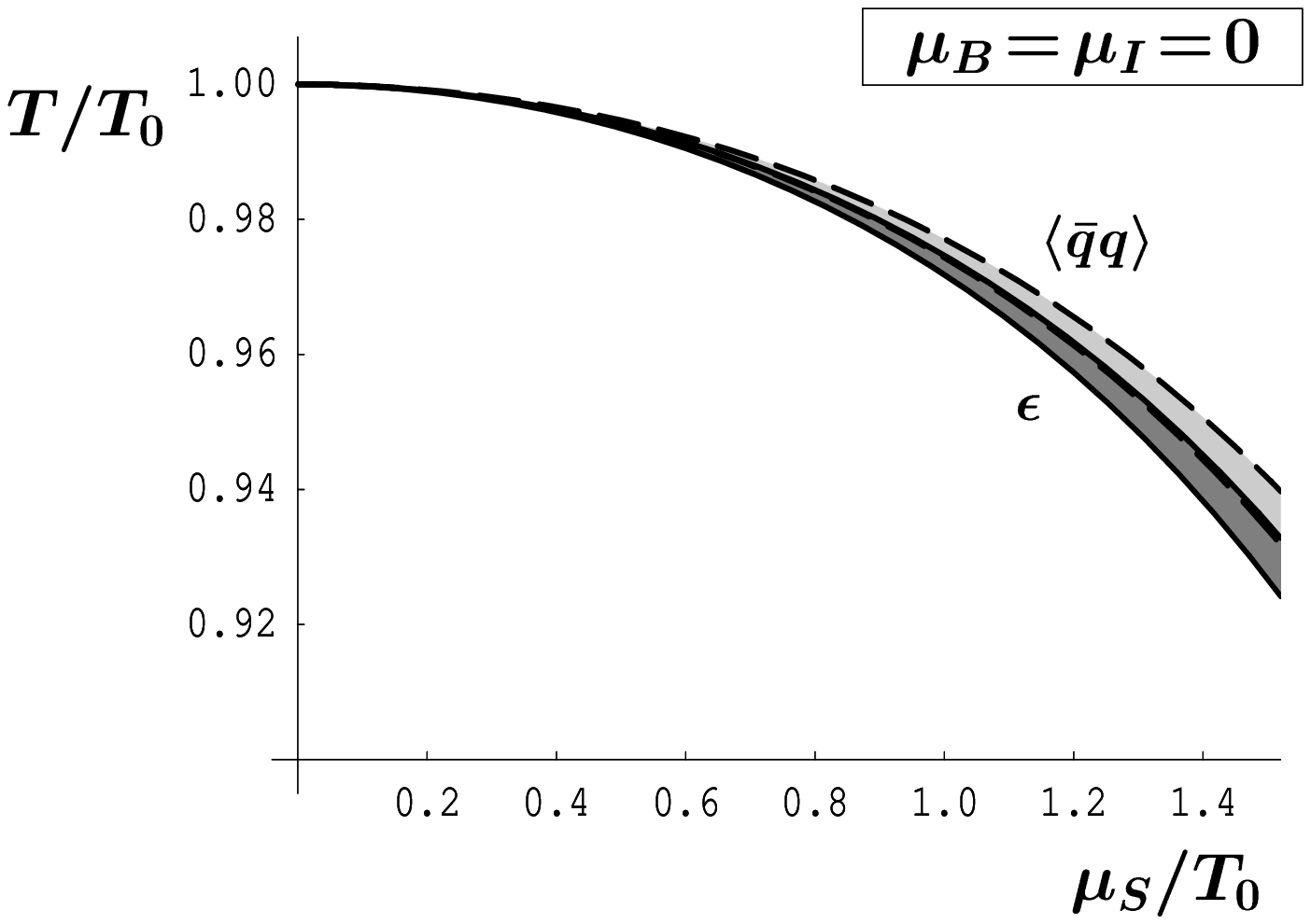}
\caption{\label{fig5} Comparison of the critical temperatures
as a function of $\mu_B$ at $\mu_I=0$ and $\mu_S=0$ (upper panel),
as a function of $\mu_I$ at $\mu_B=0$ and $\mu_S=0$
(lower left panel),
and  as a function of $\mu_S$ at $\mu_B=0$ and $\mu_I=0$
(lower right panel) obtained using the energy density
method (dark shading with full curves) and the quark-antiquark condensate
method (light shading with dashed curves). $T_0$ is the critical
temperature at zero chemical potentials.}
\end{figure}

Finally we can compare the critical temperatures obtained using these
two different approaches: constant energy density and disappearance of
the quark-antiquark condensate. We present our results in
Figure~\ref{fig5} at nonzero $\mu_B$ with $\mu_I=\mu_S=0$, nonzero
$\mu_I$ with $\mu_B=\mu_S=0$, and
nonzero $mu_S$ with $\mu_B=\mu_I=0$, respectively. 
We find that the critical temperature at zero chemical potentials, $T_0$,
is lower when we use the $\epsilon$-criterion than when we use the
$\langle\bar{q}q\rangle$-criterion. If the critical curve is
normalized with respect to $T_0$, we find that the two methods are in
very good agreement. If we compare the fits (\ref{critTenFit}) and
(\ref{critTsigmaFit}), we find that the $\mu_B$ and $\mu_S$ coefficients are
very close in both cases, whereas the $\mu_I$ coefficients
almost differ by a factor of three. However, this large difference in
the coefficients leads to critical temperatures that only differ
by a few percents in the region of interest. We therefore conclude
that both methods yield critical temperatures that are in very good
agreement.

\section{Conclusion}

We have used the hadron resonance gas model to determine the
temperature of the transition from the hadronic phase to the quark gluon
plasma phase as a function of baryon, isospin and strangeness 
chemical potentials.
This is of interest for heavy ion collision experiments, since 
baryon, isospin and strangeness 
chemical potentials are nonzero in this case.
We have used two different methods to determine the critical
temperature. The first one relies on the observation on the lattice
that the quark gluon plasma phase emerges at a constant energy
density \cite{hgrLatt}. The second method is based on the fact that the
quark-antiquark condensate for the light quarks should almost
disappear at the transition between the hadronic phase and the quark
gluon plasma phase.
We find that the critical temperatures found in both methods are in
very good agreement.

In the hadron resonance gas model, the critical temperature decreases
as the baryon, isospin, or strangeness chemical potentials increase, 
albeit slowly.
In agreement with recent lattice simulations
\cite{lattMuB_F&K,lattMuB_Bielefeld,lattMuB_ZH,lattMuB_Maria,lattMuI}
and several
models \cite{qcdMuBMuI_RMT,qcdMuBMuI_NJL,qcdMuBMuI_Ladder}, we find
that the critical temperature as a function of the quark chemical
potential at zero isospin chemical potential is almost
identical to the critical temperature as a function of isospin
chemical potential at zero quark chemical potential.
We also find that the critical temperature decreases slightly when the
isospin chemical potential is increased at fixed
baryon chemical potential. This might be important for heavy ion
collision experiments: A choice of different isotopes should reduce
the critical temperature that separates the hadronic phase from the
quark gluon plasma phase.

\begin{acknowledgments}
We thank G. Baym, S. Ejiri and K. Splittorff for useful discussions.
\end{acknowledgments}

\end{document}